\documentclass[12pt]{article}
\newlength{\vshift}
\usepackage{graphicx}
\usepackage{caption}
\usepackage{subcaption}
\usepackage{hyperref}
\newlength{\hshift}

\usepackage{amssymb}
\usepackage{amsmath}
\usepackage{amsthm}
\usepackage{amsopn}
\def\beq{\begin{equation}}
\def\eeq{\end{equation}}
\def\bea{\begin{eqnarray}}
\def\eea{\end{eqnarray}}

\title{On the Relativistic anisotropic configurations}
\author{F. Shojai$^{1,2}$, M. Kohandel$^3$,  A. Stepanian$^1$ \\ $^1$Department of Physics, University of Tehran,\\ Tehran, Iran.\\$^2$Institute for Studies in Theoretical Physics and Mathematics (IPM),\\ Tehran, Iran.\\ $^3$ ‫Department ‪of‬‬ ‫‪Physics and Chemistry,‬‬ ‫‪Alzahra‬‬ ‫‪University,\\‬‬ ‫‪Tehran, Iran‬‬.}
\date{}
\begin{document}
\maketitle
\begin{abstract}
In this paper we study anisotropic spherical polytropes within the framework of general relativity. Using the anisotropic Tolman-Oppenheimer-Volkov (TOV) equations, we explore the relativistic  anisotropic Lane-Emden equations. We find how the anisotropic pressure affects the boundary conditions of these equations. Also we argue that the behaviour of physical quantities near the center of star changes in the presence of anisotropy. For constant density, a class of exact solution is derived with the aid of a new ansatz and its physical properties are discussed.
\end{abstract}
\section{Introduction}
In theoretical astrophysics there is a growing interest to discuss the stellar structures in which the matter content is an anisotropic fluid. The effect of anisotropy can be studied both in Newtonian gravity and General Relativity. For the configurations with not extremely high densities, the presence of anisotropy factor has been discussed in Newtonian regime\cite{Herrera,Dev2,Shojai} and otherwise general relativity must be used.

 In 1972 Ruderman \cite{Ruderman} for the first time argued that the nuclear matter may have anisotropic features in very high density regime ($>10^{15} \frac{gr}{cm^3}$)  and thus the nuclear interaction needs to be treated under relativistic point of view.  After the pioneering work of Bowers and Liang \cite{Bowers} in 1974, the study of anisotropic distribution of matter got wide attention. Anisotropy arises due to the existence of solid stellar core or by the presence of type-IIIA superfluid \cite{Kippenhahn,Sokolov}, phase transitions, pion condensation in a star \cite{Sawyer}, electromagnetic field  \cite{Putney,Reimers,Martinez}, rotation, etc.

Most studies assume that the matter of the anisotropic star is described by an equation of state relating the radial pressure $P_r$ or transversal pressure $P_\perp$ to the energy density $\rho$. The polytropic equation of state is one of the most important choices for a state equation which has a wide range of applications \cite{Horedt,Nilsson,Heinzle,Kinasiewicz}. This equation of state is written as  $P_r=K\rho^\gamma$, where $\gamma=1+\frac{1}{n}$ is the polytropic exponent and $n$ is known as the polytropic index.
For $n=2$, the exact solutions of Einstein's field equations for an anisotropic sphere have been obtained by Thirukkanesh and Ragel \cite{Ragel}.

The anisotropic model has been investigated for uniform matter density in \cite{Maharaj} and for variable density distribution in  \cite{Gokhroo,Patel}.

Anisotropic pressure can affect on the stability of the stellar objects.  Dev and Gleiser \cite{Dev2,Dev1,Dev3} have extended the variational formalism of Chandrasekhar \cite{Chandra1} for isotropic star to study the stability of anisotropic general relativistic spheres against the radial perturbations. Assuming some specific density profiles and the adiabatic exponent, they have found that there can exist stable relativistic anisotropic spheres leading to instability in isotropic spheres.

Here our main purpose is to to obtain the exact solutions of general relativistic star equations in the presence of anisotropic pressure. Maharaj and Sharma \cite{Sharma} have obtained some exact solutions of compact anisotropic stars analytically by assuming a linear equation of state with an appropriate ansatz.

The investigation of anisotropic configurations was not restricted to static spherically symmetric cases. Cylinderically symmetric models have been discussed in \cite{Abbas}. 

Also the modeling of dense charged gravitating objects in strong gravitational fields has attracted much interest in recent years because of its relevance to relativistic astrophysics.  The investigation of these objects requires an exact solution of the Einstein-Maxwell system. Anisotropic fluid with a linear, quadratic and polytopic equation of state in the presence of electromagnetic field is studied in \cite{Thirukkanesh,Feroze,Takisa2,Komathiraj,Takisa}.

The importance of an equation of state in a stellar model has been emphasized by Varela et al \cite{Varela} who provided a mechanism of dealing with charged anisotropic matter in a general approach. And also the exact solutions for a charged anisotropic quark star have been reported by Sunzu et al. \cite{Sunzu}.

Anisotropy can exist not only in ordinary compact stars but also in the hypothetical objects like Boson stars \cite{Schunck} and Gravastars \cite{Visser}. Besides the isolated objects in astrophysics, anisotropy is present in globular clusters, galactic bulges and dark halos \cite{Nguyen1,Nguyen2}.

In this paper we want to discuss the anisotropic polytropes in the context of general relativity. In the next section we derive the relativistic Lane-Emden equations in the presence of anisotropy factor. Then we obtain their boundary conditions and generalize the Chandrasekhar's theorem\cite{Chandra} to the anisotropic case. In section 3, first we find a series expansion for the dimensionless Lane-Emden function, $\theta$, and the anisotropy factor, $\Delta$, near the center of star. Then in order to obtain the exact solutions, we have assumed that the presence of anisotropy factor does not change the general form of the relativistic hydrostatic equilibrium equations.
This ansatz leads us to find a new class of exact solutions for the anisotropic Lane-Emden equations for constant density. This procedure is different from the usual approach \cite{Dev2,Ragel,HerAniso} which is to suppose an ansatz on the form of anisotropy factor and try to get the physical solution of the equations. At the end in section 4, we present a summary of the main conclusions.

\section{Relativistic anisotropic polytropes}
Since we deal with the static and spherically symmetric configurations, we use the following metric:
\begin{equation}
ds^2=\exp{[\nu(r)]}dt^2-\exp[{\kappa(r)}]dr^2-r^2d\Omega^2\,,
\end{equation}
The components of energy-momentum tensor for anisotropic matter are: 
\begin{equation}T^0_0=\rho c^2 \quad ; \quad T^1_1=-P_r\quad ; \quad T^2_2=T^3_3=-P_{\perp}\end{equation}
$$T^j_k=0 \quad if \quad j\neq k$$ 
So the Einstein's field equations become: 
\begin{equation}\label{eins1}
\exp{(-\kappa)}[\frac{1}{r}\frac{d\nu}{dr}+\frac{1}{r^2}]-\frac{1}{r^2}=\frac{{8\pi G} P_ r}{c^4}
\end{equation}
\begin{equation}\label{eins2}\exp{(-\kappa)}[\frac{1}{r}\frac{d\kappa}{dr}-\frac{1}{r^2}]+\frac{1}{r^2}=\frac{{8\pi G}\rho}{c^2} \end{equation}
\begin{equation}\label{eins3}\frac{1}{2}\exp{(-\kappa)}[\frac{d^2 \nu}{dr^2}+\frac{1}{2}(\frac{d\nu}{dr})^2 +\frac{1}{r}(\frac{d\nu}{dr}-\frac{d\kappa}{dr})-\frac{1}{2}(\frac{d\nu}{dr})(\frac{d\kappa}{dr})]=\frac{{8\pi G} P_{\perp}}{c^4}\end{equation}
and the radial component of Bianchi identities reduces to the equation of hydrostatic equilibrium: 
\begin{equation} \label{TOV.rel}
\frac{dP_r}{dr}=\frac{-1}{2}\frac{d\nu}{dr}(P_r+\rho c^2)+\frac{2\Delta}{r} \quad , \quad \Delta=P_{\perp}-P_r
\end{equation}
In fact this equation can be obtained from equations (\ref{eins1}) - (\ref{eins3}) and usually it is convenient to use it instead of (\ref{eins3}). The set of equations (\ref{eins1}),(\ref{eins2}) and (\ref{TOV.rel}) contain five unknown functions, $\kappa(r)$, $\nu(r)$, $P_r(r)$, $P_{\perp}(r)$ and $\rho(r)$, and we have to add two more equations to make the set of equations closed. To do this, we use the polytropic equation of state and introduce a new ansatz for anisotropy factor which is explained in the next section. 

To obtain a unique solution of the Einstein's equations, one must also specify the boundary conditions. Outside the star, the spacetime geometry is described by the Schwarzschild metric. Thus we have: 
\begin{equation}\label{match}
\kappa=-\nu=-\ln{(1-\frac{2GM}{c^2 r})} \quad for \quad  r \geq R
\end{equation}
where $R$ is the radius of star and $M$ is its mass.
By integrating (\ref{eins2}) with initial condition $\kappa(0)=0$ (In order to avoid a singularity, $\kappa$ must tend to zero at least as  $r^2$ for $r \to 0$), we get : 
\begin{equation}
\kappa=-\ln[1-(\frac{8\pi G}{r c^2})\int_0^r\rho r^2 dr]
\end{equation}
By defining the auxiliary function: 
\begin{equation}\label{aux.rel}
u(r)=c^2 r\frac{[1-\exp{(-\kappa)}]}{2GM}=\frac{M(r)}{M} \quad ; \quad (u(0)=0 , u(R)=1)
\end{equation}
equation (\ref{eins2}) becomes: 
\begin{equation}\label{mass.eqn}
M\frac{du(r)}{dr}=4\pi\rho r^2
\end{equation}
where $M(r)$ is the relativistic mass inside the radius $r$ :
\begin{equation}\label{Mass}
M(r)=4\pi \int_0^r \rho r^2 dr
\end{equation}
By using (\ref{TOV.rel}) and (\ref{aux.rel}), we can obtain the TOV equation of hydrostatic equilibrium for the spherically static aniotropic star from (\ref{eins1}) :
\begin{equation}\label{TOV}
\frac{dP_r}{dr}=\frac{-G(\rho+\frac{P_r}{c^2})[M(r)+\frac{4\pi P_r}{c^2} r^3]}{r^2[1-\frac{2GM(r)}{c^2r}]}+\frac{2\Delta}{r}
\end{equation}
As it is obvious, vanishing $\Delta$ recovers the TOV equation for isotropic star. Moreover in the non-relativistic limit ($c\to \infty$), one can easily obtain the anisotropic Jeans equation for radial pressure:
\begin{equation}
\frac{dP_r}{dr}=\frac{-G\rho M(r)}{r^2}+\frac{2\Delta}{r} \,.
\end{equation}
The boundary conditions of (\ref{TOV}) are:
\begin{equation}
M(0)=0 \,,  P_r(0)=P_0 \,,
\end{equation}
\begin{equation}\label{BC}
M(R)=M \,, P_r(R)=0 \,.
\end{equation}
These conditions imply that the radial pressure is finite at the origin and vanishes at the surface of star. The mass function $M(r)$ goes to zero at the origin and reduces to the total mass at the surface. For a specified $\Delta$ and implying an equation of state: 
\begin{equation}\label{eos}
P_r=P_r(\rho)
\end{equation}
we obtain a closed system of equations, (\ref{Mass}), (\ref{TOV}) and (\ref{eos}). Then the metric coefficients are given by 
(\ref{TOV.rel}) and (\ref{mass.eqn}).

As we have mentioned before, for many astrophysical applications, assuming the polytropic equation of state enables one to make an approximate model of a star. One can always parametrize this equation by introducing the Lane-Emden function defined by:
\begin{equation}\label{12}
P_r=P_0\theta^{n+1} \quad , \quad \rho=\rho_0\theta^n 
\end{equation}
where $\rho_0$ and $P_0$ are the central values of relativistic density and pressure. With respect to this new function, equation (\ref{TOV.rel}) becomes: 
\begin{equation}\label{TOV.pol}
2q_0(n+1)\frac{d\theta}{dr}-\frac{4\Delta}{r\theta^n\rho_0}+(q_0\theta+1)\frac{d\nu}{dr}=0
\end{equation}
where 
\begin{equation}\label{q}
q_0=\frac{P_0}{\rho_0 c^2}\,.
\end{equation}
This equation yields: 
\begin{equation}
\nu=\nu_0+\ln{(\frac{1+q_0}{1+q_0\theta})^{2(n+1)}}-\frac{4}{\rho_0c^2}\int_0^r\frac{\Delta dr}{r\theta^n (q_0\theta+1)}
\end{equation}
where the integration constant $\nu_0$ is the central value of $\nu$. To find it,  we use the fact that  the value of the above expression is given by relation (\ref{match}) at the surface of the star . This yields to:
\begin{equation}
\nu_0=\ln{\frac{1-\frac{2GM}{Rc^2}}{(1+q_0)^{2(n+1)}}}+\frac{4}{\rho_0 c^2}\int_0^R\frac{\Delta dr}{r\theta^n (q_0\theta+1)}
\end{equation}
So we can write the metric coefficient $\nu(r)$  in terms of $\theta$ and $\Delta$ as:  
\begin{equation}
\nu(r)=\ln{\frac{1-\frac{2GM}{Rc^2}}{(1+q_0\theta)^{2(n+1)}}}+\frac{4}{\rho_0 c^2}\int_r^R\frac{\Delta dr}{r\theta^n (q_0\theta+1)}
\end{equation}
Substituting $e^{-\kappa}$ from (\ref{aux.rel}) and $\frac{d\nu}{dr}$ from (\ref{TOV.pol}) into equation (\ref{eins1}), and using (\ref{mass.eqn}), the above equation reduces to:
\begin{multline}\label{23}
\frac{q_0 (n+1)\frac{d\theta}{dr} r}{(1+q_0\theta)}(1-\frac{2GM}{rc^2}u)+\frac{GMu}{rc^2}+\frac{GMq_0\theta}{c^2}\frac{du}{dr}+\\ 
\frac{2\Delta}{\rho_0 c^2 \theta^n (1+q_0\theta)}(1-\frac{2GMu}{rc^2})=0
\end{multline}

Now defining the dimensionless parameters $\xi$ and $\eta$ : 
\begin{equation}\label{dimensionless}
\eta=\frac{M}{4\pi \rho_0 \alpha^3}u \quad \,, \quad r=\alpha\xi
\end{equation}
where \footnote{The plus sign holds if $-1<n<\infty$, and the minus sign if $-\infty<n<-1$}
$\alpha^2=\pm \frac{(n+1)P_0}{4\pi G \rho_0^2}$ 
, equations (\ref{mass.eqn}) and (\ref{23}) take the following form: 
\begin{equation}\label{L.E.n1}
\frac{\pm \xi-2q_0(n+1){\eta}}{1+q_0\theta}(\xi\frac{d\theta}{d\xi}+2\frac{\Delta}{\rho_0 c^2 \theta^n (n+1)q_0})+\eta+q_0\theta\xi \frac{d\eta}{d\xi}=0
\end{equation}
\begin{equation}\label{L.E.n2}
\frac{d\eta}{d\xi}=\theta^n\xi^2
\end{equation}
which are the general relativistic Lane-Emden equations for anisotropic polytropes. 
In the non-relativistic  limit, according to (\ref{q}), $q_0 \sim 0$ and one can recover the Newtonian anisotropic Lane-Emden equation \cite{Shojai}. Also setting $\Delta=0$, the isotropic Lane-Emden equation is obtained.

In equation (\ref{L.E.n1}) if $\xi $ tends to zero, we will have: 
\begin{equation}
\frac{d\theta}{d\xi}|_{\xi \to 0}=\frac{\mp 2\Delta(0)}{\rho_0c^2(n+1)q_0 \xi}|_{\xi \to 0}
\end{equation}
 So we arrive at the following theorem:

\textbf{Theorem} :\textit{The finite solutions of relativistic anisotropic Lane-Emden equations at the origin have to satisfy the following relations}:\begin{equation}\label{init}\frac{d\theta}{d\xi}|_{\xi \to 0}=\frac{\mp 2\Delta(0)}{\rho_0c^2(n+1)q_0 \xi}|_{\xi \to 0} \quad \,, \quad \theta(0)=1 \quad \,, \quad \eta(0)=0
\end{equation}
This is the generalization of theorem 1 in \cite{Shojai} for the relativistic anisotropic case. 
\section{Analytical solutions of relativistic Lane-Emden equation}
In this section we shall discuss how one can solve the Lane-Emden equations (\ref{L.E.n1}), (\ref{L.E.n2}) with the boundary conditions (\ref{init}). These are a system of two equations for three unknowns, the function $\theta(\xi)$, $\Delta(\xi)$ and $\eta(\xi)$. To solve these equations, we need another equation. One can assume either a special form for one of the above functions or a mathemtaical relation between these functions in order to the set of equations be closed. Here we use the latter approach which is explained in Section 3.2.

First in Section 3.1 we find a series expansion of $\theta(\xi)$, $\Delta(\xi)$ and $\eta(\xi)$ near the center of anisotropic star. For isotropic case one can find the corresponding series expansion in\cite{Horedt} and the references in it. That is: 
\begin{equation}\theta \approx 1+ a \xi^2 + b \xi^4+...\end{equation}
\begin{equation}\eta=\frac{\xi^3}{3}+na\frac{\xi^5}{5}+\frac{[nb+n(n-1)\frac{a^2}{2}]\xi^7}{7} \end{equation}
where
\begin{equation}\label{coefi1}
a=\mp\frac{(\frac{1}{3}+4\frac{q_0}{3}+q_0^2)}{2}  
\end{equation}
\begin{equation}\label{coefi2}
 \quad b=[\frac{n}{15}+\frac{2nq_0}{9}+(\frac{16n}{45}+\frac{2}{3})q_0^2+(\frac{6n}{5}+\frac{8}{3})q_0^3+(n+2)q_0^4]/8 \,.
\end{equation}

 In Section 3.2 we shall solve the Lane-Emden equations assuming the density is constant throughout the star. A similar calculation for isotropic star has been derived by Tooper \cite{Tooper}. 
In this case, the variables $\theta$ and $\eta$ are separable and the exact analytical solutions of (\ref{L.E.n1}) and (\ref{L.E.n2}) for $\Delta=0$, are \cite{Tooper}: 
\begin{equation}\label{solI}
\theta=\frac{[(1+3q_0)\sqrt{1-\frac{2q_0\xi^2}{3}}-(1+q_0)]}{q_0[3(1+q_0)-(1+3q_0)\sqrt{1-\frac{2q_0\xi^2}{3}}]}
\end{equation}
\begin{equation}\eta=\frac{\xi^3}{3}\,.\end{equation}
\subsection{Series expansion}
Near the center of star, we assume that:
$$\theta(\xi) = \sum_{k=0}^{\infty} a_k \xi^k$$
$$\Delta(\xi) = \sum_{k=0}^{\infty} b_k \xi^k$$
\begin{equation}\label{approx}
\eta(\xi) = \sum_{k=0}^{\infty} d_k \xi^k
\end{equation}
Inserting these expressions into the Lane-Emden equations  (\ref{L.E.n1}) and (\ref{L.E.n2}), we get:
\begin{equation}\label{coef1}
a_0=1, \quad  a_1=\frac{\pm2 \Delta'(0)}{\rho c^2 (n+1)q_0 }, \quad a_2=\mp \frac{1}{2}(\frac{1}{3}+q_0^2+\frac{4}{3}q_0)
\end{equation}
\begin{equation} \label{coef2}
 b_0=0, \quad b_1=\Delta'(0) 
\end{equation}
\begin{equation} \label{coef3}
d_1=0, \quad  d_2=0, \quad d_3=\frac{1}{3}, \quad d_4=\frac{\mp 2n \Delta'(0)}{4c^2(n+1)q_0}\,. \end{equation}
provided that the pressure and density gradient (and also $\Delta(0)$ according to equation (\ref{init}) ) are finite at the center of star. Thus if $r\to 0$, $\eta \approx \xi^3$, $\Delta \approx \xi$ and $\theta \approx 1$. Comparing equations (\ref{coef1})-(\ref{coef3}) with (\ref{coefi1}) and (\ref{coefi2}), one can find that how the anisotropy factor affects the behaviour of quantities near the center of star.
\subsection{An exact solution}
Following \cite {Shojai}, we assume that the presence of $\Delta$ does not change the forms of isotropic Lane-Emden equations  and it only modifies the coefficients of these equations. To do this, noting equation (\ref{TOV.pol}), 
we introduce a constant $m$ such that (hereafter we set $c=1$): 
\begin{equation}
2q_0 (n+1) d\theta-\frac{4\Delta}{r\rho_0{\theta}^n}dr=2q_0(m+1)d\theta
\end{equation}

Substituting this into equation (\ref{TOV.pol}) and integrating it, the metric coefficient $\nu(r)$ becomes: 
\begin{equation}\label{15}
\nu(r)=\ln[\frac{(1-\frac{2GM}{R})}{(1+q_0\theta)^{2(m+1)}}] 
\end{equation}
which is identical to the isotropic one \cite{Horedt} in which the  coefficient $n$ has changed to $m$. Using equations (\ref{aux.rel}) and (\ref{15}) and defining a dimensionless parameter $\xi$ such that: 
\begin{equation}\label{alfa}
r=\tilde{\alpha}\xi \qquad  (m\neq -1,\pm \infty)
\end{equation}
where $ \tilde{\alpha}=[\pm \frac{(m+1)P_0}{4\pi G \rho_0^2}]^{\frac{1}{2}}$.
Equations (\ref{L.E.n1}) and (\ref{L.E.n2}) reduce to: 
\begin{equation}\label{LE1}
\{ \frac{[\pm1 - 2q_0(m+1)\frac{\eta}{\xi}]}{1+q_0 \theta}\}\xi^2 \frac{d\theta}{d\xi}+\eta + q_0\xi \theta \frac{d\eta}{d\xi}=0
\end{equation}
\begin{equation}\label{LE2}
\frac{d\eta}{d\xi}=\theta^n\xi^2. 
\end{equation}
Equations (\ref{LE1}) and (\ref{LE2}) are the extended form of relativistic Lane-Emden equations for anisotropic configurations, resulted by our ansatz describing above.

 Here we want to obtain the analytical solution of anisotropic Lane-Emden equations for constant density. For $n=0$, a solution of equations (\ref{LE1}) and (\ref{LE2})  satisfying the initial conditions (\ref{init}) is: 

\begin{equation}\label {Sol}
\theta=\frac{(1+3q_0){(1-\frac{q_0\xi^2}{2k})}^{\frac{2k}{3}}-(1+q_0)}{q_0[3(1+q_0)-(1+3q_0){(1-\frac{q_0\xi^2}{2k})}^{\frac{2k}{3}}]}
\end{equation}
in which $k=\frac{3}{4(m+1)}$ is a natural number and $q_0 > 0$. This solution vanishes at: \begin{equation}\label{zerodel}\xi_R=\sqrt{\frac{2k}{q_0}[1-(\frac{q_0+1}{3q_0+1})^{\frac{3}{2k}}]}\end{equation} which is the dimensionless value of the radial coordinate at the boundary. Thus we find that:
\begin{equation}\label{Sol delta}
\frac{\Delta}{P_0}= \frac{(1+q_0)(1+3 q_0) (3-4 k) \xi^2 \left(1-\frac{q_0 \xi^2}{2 k}\right)^{-1+\frac{2 k}{3}} }{6 k \left(-3 (1+q_0)+(1+3 q_0) \left(1-\frac{q_0 \xi^2}{2 k}\right)^{2 k/3}\right)^2}
\end{equation}
With suitable choices of $k$ and $q_0$, the equations of (\ref{Sol}) and (\ref{Sol delta}) present a new class of solution for an anisotropic constant-density star.

\begin{figure}[h]
\begin{center}
\fbox{\includegraphics[scale=0.8]{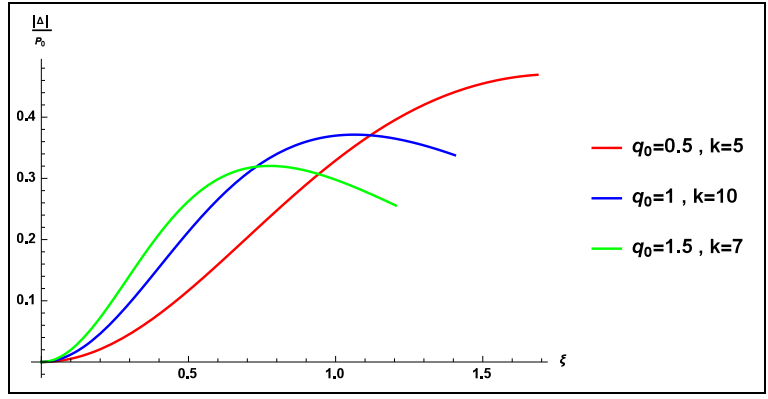}}
\caption{Plot of anisotropy factor}{\label{fig1}}
\end{center}
\end{figure}

\begin{figure}[h]
\begin{center}
\fbox{\includegraphics[scale=0.8]{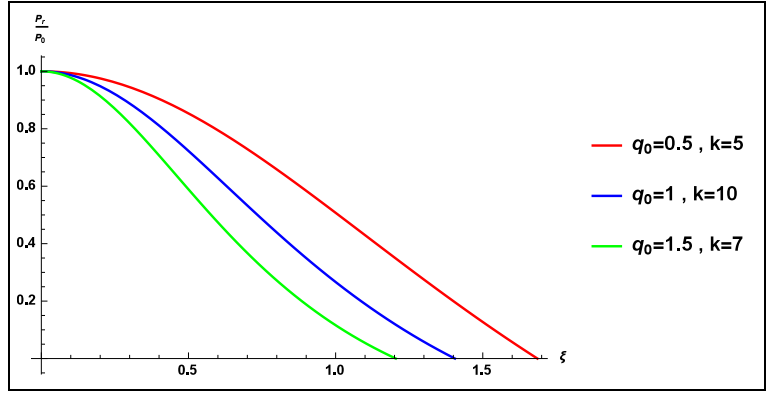}}
\caption{Plot of radial pressure}{\label{fig2}}
\end{center}
\end{figure}

This solution can be considered as a well-defined solution if the physical quantities like the anisotropy factor be finite inside the star. It can be easily shown that for all values of $q_0$ and $k$, the anisotropy factor  (\ref{Sol delta}) is finite. Also if $q_0\ge 1$, the anisotropy factor has a local maximum inside the star but this is not true for $q_0 < 1$. Here in Figure \ref{fig1} we have plotted $\frac{\mid \Delta \mid}{P_0}$ as a function of $\xi$ for some values of $q_0$ and $k$. For any choice, the plots are from the center to the corresponding radius of the star.
The radial pressure is plotted in Figure \ref{fig2} for the same values of $k$ and $q_0$ as in Figure \ref{fig1}. Now it becomes clear that it is a positive regular function everywhere inside the star. It starts with its local maximum value at the center of the star and gradually decreases until it vanishes at the surface of star. 

\newpage

\begin{figure}[h]
\begin{center}
\fbox{\includegraphics[scale=0.8]{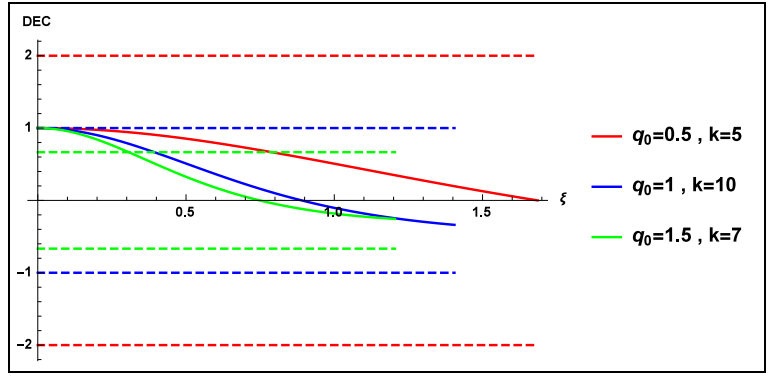}}
\caption{Plot of the dominant energy condition} {\label{fig3}}
\end{center}
\end{figure}

 Now let us to discuss the validity of energy conditions. Since $q_0$ is positive, the weak, null and strong energy conditions hold for every positive value of central pressure. A simple calculation shows that for cases with $0 < q_0 \le 1$,  the dominant energy condition is satisfied same as the other conditions. But for cases with $q_0 > 1$, this condition is not satisfied within  a radius of $\sqrt{\frac{2k}{q_0}\{1-\frac{1}{2}[(1+\frac{2}{3}k)\sqrt{\frac{2(1+q_0)}{1+3q_0}}]^{\frac{3}{k}}\}}$ from the center of star.
In Figure \ref{fig3}, we have plotted the dominant energy condition as a function of $\xi$ for different values of $q_0$ and $k$. While the dashed lines in each cases represent $-\frac{1}{q_0}$ and $\frac{1}{q_0}$, it is clear that for $q_0=\frac{1}{2}$ there is a violation of the dominant energy condition. 
\newpage
Other quantities of physical interest are the stellar radius and mass. To find these, it is needed to find the radius of the star ,$R'$, measured by an external observer. This radius for an anisotropic polytropic sphere with constant density is:
\begin{multline}\label{Radius}
R'=\int_0^R \exp[\frac{\kappa(r)}{2}]dr=\tilde{\alpha} \int_0^{\xi_R} \frac{d\xi}{\sqrt{1-2q_0(m+1)\frac{\xi^2}{3}}}= \\ \tilde{\alpha} (\frac{3}{2q_0(m+1)})^{1/2}\arcsin{(\frac{2q_0(m+1)\xi_R^2}{3})^{1/2}}
\end{multline}
For $m=0$, relation (\ref{Radius}) reduces to the corresponding radius in the case of isotropic star\cite{Horedt}.
Now one can find the mass-radius relationship from (\ref{Mass}):
\begin{equation}\label{47}
M=4\pi \rho_0\tilde{\alpha}^3 \int_0^{\xi_R} \xi^2 d\xi=\frac{4\pi \rho_0\tilde{\alpha}^3 \xi_R^3}{3}
\end{equation}
then according to relation (\ref{alfa}) :
\begin{equation}
M=\frac{(m+1)q_0 \xi_R^2 R}{3G}
\end{equation}
and using the value of $\xi_R$ (\ref{zerodel}), one can find the mass of an anisotropic star: 
\begin{equation}
M=\frac{1}{3}(\frac{G}{4\pi \rho_0})^{\frac{1}{2}}\{\frac{3}{2}[1-(\frac{q_0+1}{3q_0+1})^{\frac{3}{2k}}]\}^{\frac{3}{2}} \,.
\end{equation}
To get the mass-radius relationship, it is required to evaluate the integrals of (\ref{Radius}) and (\ref{47}) with arbitrary radial coordinate which now appears as the upper limit of integrals. We thus get:
\begin{equation}\label{MassRadius}
\frac{M(r)}{r'}=\sqrt{\frac{128 (\rho_0 \pi G)^3}{27}}\frac{r^3}{\arcsin{[\frac{8}{3}\pi G \rho_0 r^2]^{1/2}}} \,.
\end{equation}

\begin{figure}[h]
\begin{center}
\fbox{\includegraphics[scale=0.8]{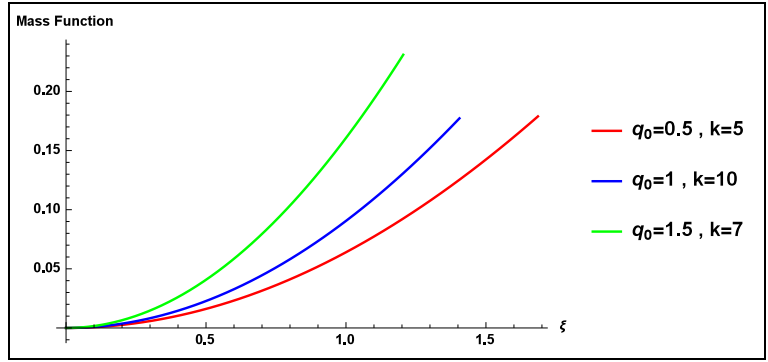}}
\caption{Plot of mass-radius relation} {\label{fig4}}
\end{center}
\end{figure}

Here in order to plot the mass-radius relation we replace both $M(r)$ and $r$ with their dimensionless parameters $\eta(\xi)$ and $\xi$. Figure \ref{fig4} shows the dimensionless form of the mass-radius relation (\ref{MassRadius}) as a function of $\xi$ for specified values of $k$ and $q_0$.

Also the internal compactness of the star can be written as follows: 
\begin{equation}
\frac{M(r)}{r}=\sqrt{\frac{(4\pi G \rho_0)^3}{q_0(m+1)}}r^2
\end{equation}
and correspondingly the surface redshift $z(r)$ is given by : 
\begin{equation}
1+z(r)=[1-2\frac{M(r)}{r}]^{-1/2}=[1-2\sqrt{\frac{(4\pi G \rho_0)^3}{q_0(m+1)}}r^2]^{-1/2} \,.
\end{equation}
\newpage
\section{Concluding remarks}
In this paper, we have discussed the relativistic Lane-Emden equations and their boundary conditions for an anisotropic star. We made a new proposal to find the exact solution of these equations for constant density. Noting equation (\ref{TOV.pol}), we assume that the effect of anisotropy is changing the coefficient of the first term from $2 q_0(n+1)$ to $2 q_0(m+1)$ or alternatively changing the coefficient of the third term from $(q_0\theta+1)$ to $m(q_0\theta+1)$ where $m$ is different from polytropic index in general. The first case is studied in section (3.2) with details. For the second case using equation (\ref{TOV.pol}), one can introduce $m$ such that: 
\begin{equation}
\frac{m}{2}[P_0\theta+\rho_0]\frac{d\nu}{dr}=\frac{1}{2}\frac{d\nu}{dr}(P_0\theta+\rho_0)+\frac{-2\Delta}{r}
\end{equation}

So the resulted Lane-Emden equations will be :
\begin{equation}
\{\frac{[\pm1 -\frac{2q_0(n+1)}{m}\frac{\eta}{\xi}]}{1+q_0\theta}\}\xi^2 \frac{d\theta}{d\xi}+\eta+q_0\xi \theta \frac{d\eta}{d\xi}=0 \quad ,\frac{d\eta}{d\xi}=\xi^2 \theta^n
\end{equation}
where the plus sign holds if $-1<\frac{n+1}{m}<\infty$ and the minus sign if $-\infty<\frac{n+1}{m}<-1$.
In this case the dimensionless parameter $\xi$ is introduced as $r$=$(\frac{\pm \frac {n+1}{m} q_0}{4\pi G \rho_0 })^{1/2}{\xi}$.
These equations for $n=0$ yield the same solution (\ref{Sol}) and (\ref{Sol delta}) in which  $k=\frac{3}{4}m$. 
\par
{\bf Acknowledgments:}
This work was supported by a grant from University of Tehran.

\end{document}